%% file: 0_main.tex
\input{0_headers}
\usepackage[utf8]{inputenc}

\newcommand{\new}{\vspace{0.07in}}
\renewcommand{\hl}[1]{#1}					

\begin{document}
\maketitle

\input{abstract_intro}

\input{rest_of_paper}

\bibliography{aaai24.bib}
\input{appendix}
\end{document}

%% file: 0_headers.tex
\documentclass[letterpaper]{article} 
\usepackage{aaai24}
\usepackage{times}  
\usepackage{helvet}  
\usepackage{courier}  
\usepackage[hyphens]{url}  
\usepackage{graphicx} 
\usepackage{natbib}  
\usepackage{caption} 
\usepackage{soul, color}
\usepackage{fancyhdr}
\usepackage{algorithm}
\usepackage{algorithmic}
\usepackage{amsmath}
\usepackage{newfloat}
\usepackage{listings}
\usepackage{multirow}
\usepackage{amssymb}
\usepackage[table]{xcolor}
\DeclareCaptionStyle{ruled}{labelfont=normalfont,labelsep=colon,strut=off} 
\floatstyle{ruled}
\newfloat{listing}{tb}{lst}{}
\floatname{listing}{Listing}
\DeclareMathOperator*{\argmin}{argmin}
\DeclareMathOperator*{\argmax}{argmax}

\newcommand{\name}{{\sc OracleBO}}
\title{Sample-Constrained Black Box Optimization for Audio Personalization}
\author{Rajalaxmi Rajagopalan, Yu-Lin Wei, Romit Roy Choudhury\\
}
\affiliations {
    Department of Electrical \& Computer Engineering\\
University of Illinois at Urbana-Champaign\\
    \{rr30,yulinlw2,croy\}@illinois.edu
}
\vspace{-7.5ex}

\usepackage{bibentry}

%% file: abstract_intro.tex
\begin{abstract}
We consider the problem of personalizing audio to maximize user experience.
Briefly, we aim to find a filter $h^*$, which applied to any music or speech, will maximize the user's satisfaction.
This is a black-box optimization problem since the user's satisfaction function is unknown.
Substantive work has been done on this topic where the key idea is to play audio samples to the user, each shaped by a different filter $h_i$, and {\em query} the user for their satisfaction scores $f(h_i)$.
A family of ``surrogate'' functions is then designed to fit these scores and the optimization method gradually refines these functions to arrive at the filter $\hat{h}^*$ that maximizes satisfaction.

In certain applications, we observe that a second type of querying is possible where users can tell us the individual elements $h^*[j]$ of the optimal filter $h^*$.
Consider an analogy from cooking where the goal is to cook a recipe that maximizes user satisfaction. 
A user can be asked to score various cooked recipes (e.g., tofu fried rice) or to score individual ingredients (say, salt, sugar, rice, chicken, etc.). 
Given a budget of $B$ queries, where a query can be of either type, our goal is to find the recipe that will maximize this user's satisfaction.

Our proposal builds on Sparse Gaussian Process Regression (GPR) and shows how a hybrid approach can outperform any one type of querying.
Our results are validated through simulations and real world experiments, where volunteers gave feedback on music/speech audio and were able to achieve high satisfaction levels. 
We believe this idea of hybrid querying opens new problems in black-box optimization and solutions can benefit other applications beyond audio personalization.

\end{abstract}

\vspace{-1em}
\section{Introduction}
Consider the problem of personalizing content to a user's taste. 
Content could be audio signals in a hearing aid, a salad cooked for the user, a personalized vacation package designed by an AI agent, etc.
Given the content $c$, we intend to adjust the content with a linear filter $h$. 
Our goal is to find the optimal filter $h^*$ that will maximize the user's personal satisfaction $f(h)$.
\new 

Finding $h^*$ is difficult because the user's satisfaction function $f(h)$ is unknown; it is embedded somewhere inside the perceptual regions of the brain.
Hence, gradient descent is not possible since gradients cannot be computed. 
Black box optimization (BBO) has been proposed for such settings, where one queries user-satisfaction scores for carefully chosen filters $h_i$.
Using a budget of $B$ such queries, BBO expects to estimate $\hat{h}^*$ that is close to the true $h^*$. 
Of course, reducing the query budget $B$ is of interest, and the effects of lowering $B$ have been studied extensively.
\new

The above problem can be called ``{\em filter querying}'' because the user is queried using different filters $h_i \in \mathbf{R}^N$.
In this paper, we discuss an extension to this problem where a second type of querying is possible, called ``{\em dimension querying}''.
With dimension querying, it is possible to query the user for each dimension of the optimal filter, namely $h^*[1], h^*[2], \dots, h^*[N]$.
In audio personalization, for example, the optimal $h^*$ is the hearing profile of a user in the frequency domain; if we accurately estimate the hearing profile, we can maximize their satisfaction.
With dimension querying, a user can listen to sound at each individual frequency $j \in \{1,N\}$ and tell us the best score $h^*[j]$.
The only problem is that $N$ can be very large, say $8000$ Hz, hence it is prohibitive to query the user thousands of times.
\new 

To summarize, a filter query offers the advantage of understanding user satisfaction for a complete filter, while a dimension query gives optimal information for only one dimension at a time. 
In our analogy from cooking, filter querying gives us the user satisfaction for a fully prepared recipe (e.g., tofu fried rice), while dimension querying gives us the user's optimal liking for individual dimensions, like salt, sugar, etc.
This paper asks, for a given query budget $B$, can the combination of two types of querying lead to higher user satisfaction compared to a single type of querying?
How much is the gain from the combination, and how does the gain vary against various application parameters?
\new 

Our solution builds on past work that uses Gaussian Process Regression (GPR).
Conventional GPR begins with a family of surrogate functions for $f(h)$ and estimates a posterior on these functions, based on how well they satisfy the user's scores.
Over time, GPR iterates through a two-step process, first picking an optimal filter $h_i$ to query the user, and based on the score $f(h_i)$, updates the posterior distribution.
The goal is to query and update the posterior until a desired criteria is met. 
Once met, the mean of this posterior $\hat{f}(h)$ is declared as the surrogate for $f(h)$ and $\hat{h}^* = \argmax \hat{f}(h)$ is announced as the final personalization filter.
\new 

When applied to our cooking analogy, conventional GPR will prescribe different recipes and based on scores from the user, will construct a posterior on candidate satisfaction functions.
The next recipe will be chosen to be one that could return a higher score than all past scores (say,  Japanese sushi).
In essence, GPR tries to reason about the shape of the satisfaction function using only recipe-based querying. 
\new

Our contribution lies in {\em querying the user with better recipes} to expedite the process of estimating the satisfaction function.
We modify the GPR framework to first estimate a batch of $q$ filters (instead of just one) from the posterior; we then choose a single winner based on which has the strongest similarity to the dimension-scores.
Finally, these operations are performed after GPR has been transformed into a sparse space, otherwise it becomes difficult to meet the query budget $B$. 
\new

Results show that spending some query budget on dimension queries (salt and sugar) as opposed to spending all the budget on filter queries (full recipes) offers benefits.
We lack mathematical proof, instead show empirical results from extensive simulations and real-world experiments.
With real volunteers who were asked to rate audio quality on a scale of $[0 - 10]$, our proposed method, {\name}, achieves an average of $3.3$ points higher satisfaction, within a budget of $B=30$ queries.
Using simulations with various synthetic satisfaction functions, we find that the break-down between the two types of queries exhibits a sweet spot.
When the number of dimension queries increases (or decreases) beyond a fraction of $B$, the achieved maxima deviates from the global maxima.
We present sensitivity analysis and ablation studies, and discuss a number of follow-up questions for future research.

%% file: rest_of_paper.tex
\section{Problem Formulation}\label{sec:problem}
Consider an {\em unknown} real-valued function $f: \mathcal{H} \to \mathbf{R}$ where the domain is a compact subspace $\mathcal{H} \subseteq \mathbf{R}^N$, $N \ge 500$. Let $h^*$ be the minimizer of $f(h)$. 
We want to estimate $h^*$ using a budget of $B$ queries, where a query can be one of two types:

\begin{itemize}
\item[1.] The unknown function $f$ can be sampled at a given $h_i$. This query yields $f(h_i)$. We call these {\em filter queries}, $Q_f$.

\item[2.] An Oracle is assumed to know information about the minimizer $h^*$. The Oracle when queried can  give us one dimension of the vector $h^*$, i.e., we can obtain $h^*[j]$, for any given $j \in [1,2,\dots,N]$. We call these {\em dimension queries}, $Q_d$.
\end{itemize}

Thus, the optimization problem is shown in Eqn. \ref{eqn:opti-prob}.
\begin{equation}
\begin{aligned}
\argmin_{\hat{h} \in \mathcal{H}} \quad &||f(\hat{h})-f(h^*)||_2
\\
\textrm{s.t.} \quad &Q_f + Q_d \leq B
\end{aligned}
\label{eqn:opti-prob}
\end{equation}
where $Q_f, Q_d$ are the number of filter and dimension queries described above, and the query budget $B \ll N$.
\new

The unknown function $f$ may be non-convex, may not have a closed-form expression, and its gradient is unavailable.
However, since it can be queried for a given $h_i$, it is called a black-box optimization problem.
We approach this through Bayesian Optimization that builds on Gaussian Process Regression (GPR), Expected Improvement (EI) acquisition function, and sparsity transformations (to cope with the large gap between $N$ and $B$).
We review the relevant background on Bayesian optimization next, followed by our proposed algorithm, {\name}.
\section{Bayesian Optimization}
Bayesian optimization \cite{frazier2018tutorial} broadly consists of the following two modules:

\begin{itemize}
\item[(1)] \textbf{Surrogate model}: 
A family of functions that serve as candidates for the unknown objective function. 
The functions are commonly drawn from a Gaussian process generated by \textbf{Gaussian Process Regression (GPR)}. 
This essentially means that GPR generates a Gaussian posterior distribution of the likely values function $f$ can take at any point of interest $h$.

\item[(2)] \textbf{Acquisition function}: A sampling strategy that prescribes the point at which $f$ should be observed next. 
The GPR posterior model is used to evaluate the function at new points $h'$, and one is picked that maximizes a desired metric. 
This new point $h'$ when observed will \hl{maximally improve the GPR posterior.} 
\end{itemize}

We review GPR next, followed by a popular acquisition function called ``Expected Improvement''.

\subsection{Gaussian Process Regression (GPR)}

\begin{figure}[htbp]
\centerline{\includegraphics[clip, width=1.0\columnwidth]{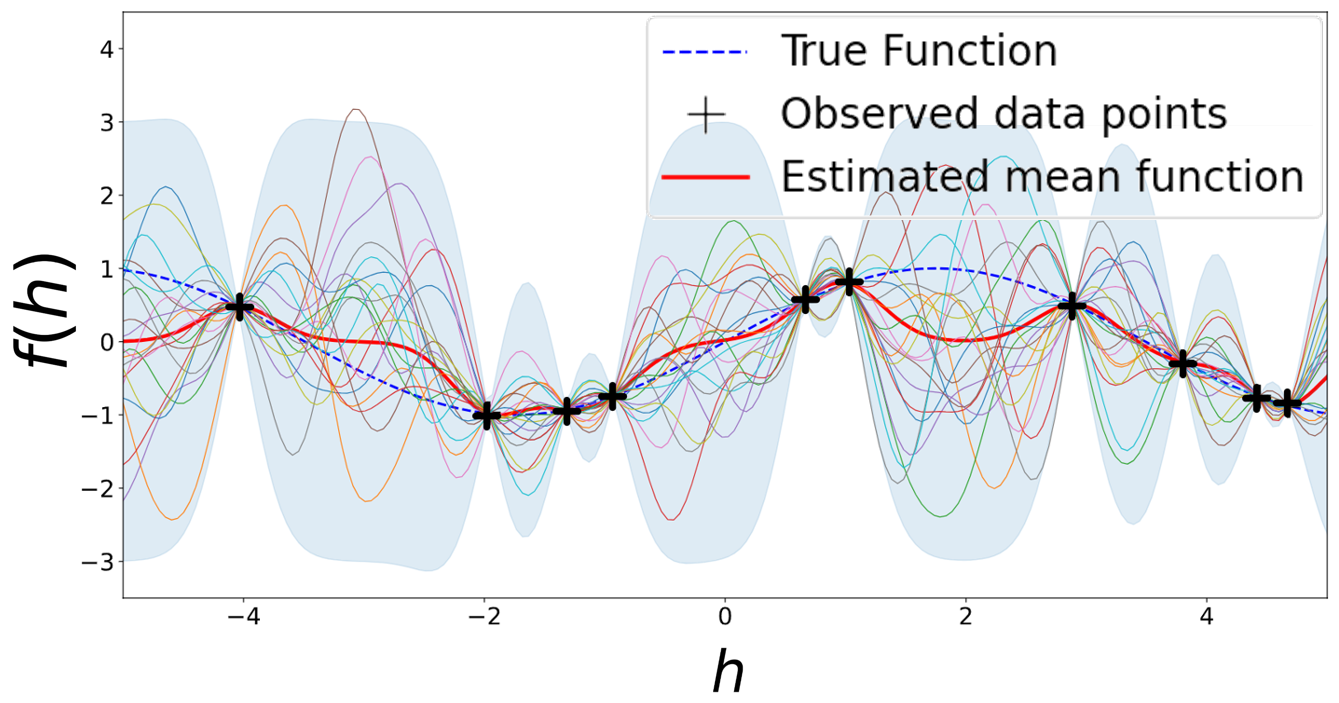}}
\caption{\small{A GPR Posterior: 
The black ``plus" symbols mark all the observed points.  
The dashed-blue line is the true $f$, and the red line is the estimated $\hat{f}$ (the mean of the posterior). 
The light blue shaded area marks the variance.}}
\label{fig:posterior}
\end{figure}

\textbf{Non-parametric Model}: 
Gaussian processes \cite{wang2020intuitive} are helpful for black-box optimization because they provide a non-parametric mechanism to generate a surrogate for the unknown function $f$. 
Given a set of samples $\mathcal{X} = \{h_1,h_2,\dots,h_K\}$ at which the function $f$ has been observed, i.e., we know $\mathcal{F} = \{f(h_1),f(h_2),\dots,f(h_K)\}$, we can identify an infinite number of candidate functions that match the observed function values. 
Figure \ref{fig:posterior} shows an example function $f$ in 1-dimensional space. 

\textbf{Function Distribution \& Kernel}: 
GPR generates the surrogate model by defining a Gaussian distribution over these infinite candidate functions. 
Given the set of observations $(\mathcal{X},\mathcal{F})$, the mean $\boldsymbol\mu$ of the distribution is the most likely surrogate of the function $f$.
The covariance $\mathbf{K}$ is a kernel that dictates the smoothness (or shape) of the candidate functions and must be chosen based on domain knowledge of $f$. 
One commonly used kernel is the ARD (Automatic Relevance Determination) {\em power exponential kernel}:
\begin{equation}
\begin{aligned}
k(h,h') = a_0\exp{-\frac{1}{2}(h-h')^T\Sigma^{-1}(h-h')}
\end{aligned}
\label{eqn:kernel}
\end{equation}
where, $a_0$ and $\Sigma = \text{diag}_i(\sigma_i)$ are the kernel parameters. 
\new

\textbf{Prior \& Posterior}: 
Before any observations, the distribution defined by $\mathbf{K}$ and $\boldsymbol\mu = \mathbf{0}$ forms the prior distribution. Given a set of observations $(\mathcal{X},\mathcal{F})$, the prior is updated to form the posterior distribution over the candidate functions. Figure \ref{fig:posterior} shows the distribution of candidates and the mean surrogate model of an example $f$.
With more observations, the current posterior serves as the prior, and the new posterior updates from the new observations. 
Eqn. \ref{eqn:posterior} models the function $f$ with the posterior generated by GPR.

\begin{equation}
\begin{aligned}
P(\mathcal{F}|\mathcal{X}) \sim \mathcal{N}(\mathcal{F}|\boldsymbol\mu,\mathbf{K})
\end{aligned}
\label{eqn:posterior}
\end{equation}

where, $\boldsymbol\mu$ = $\{\mu(h_1),\mu(h_2),\dots,\mu(h_K)\}$ and $\mathbf{K}_{ij}$=$k(h_i,h_j)$, $k$ represents a kernel function. 
\new

\textbf{Predictions}: To make predictions $\hat{\mathcal{F}}=f(\hat{\mathcal{X}})$ at new points $\hat{\mathcal{X}}$, GPR uses the current posterior $P(\mathcal{F}|\mathcal{X})$ to define the joint distribution of $\mathcal{F}$ and $\hat{\mathcal{F}}$, $P(\mathcal{F},\hat{\mathcal{F}}|\mathcal{X},\hat{\mathcal{X}})$ in Eqn. \ref{eqn:predict}.

\begin{equation}
\begin{aligned}
\begin{bmatrix}
    \mathcal{F} \\ \hat{\mathcal{F}}
\end{bmatrix} \sim \mathcal{N}\left(\begin{bmatrix}
    \mu(\mathcal{X}) \\ \mu(\hat{\mathcal{X}})
\end{bmatrix}, \begin{bmatrix}
    \mathbf{K} & \hat{\mathbf{K}} \\ \hat{\mathbf{K}}^T & \hat{\hat{\mathbf{K}}}
\end{bmatrix}\right)
\end{aligned}
\label{eqn:predict}
\end{equation}

where, $\mathbf{K} = k(\mathcal{X},\mathcal{X})$, $\hat{\mathbf{K}} = k(\mathcal{X},\hat{\mathcal{X}})$, $\hat{\hat{\mathbf{K}}} = k(\hat{\mathcal{X}},\hat{\mathcal{X}})$ and $(\mu(\mathcal{X}), \mu(\hat{\mathcal{X}})) = \mathbf{0}$. 
\new

The conditional distribution and hence prediction of $\hat{\mathcal{F}}$ is derived from the joint distribution shown in Eqn. \ref{eqn:predict-condt}. 
The proof and explanations of all the above are clearly presented in \cite{wang2020intuitive}).
\vspace{-0.75em} 
\begin{equation}
\begin{aligned}
P(\hat{\mathcal{F}}|\mathcal{F},\mathcal{X},\hat{\mathcal{X}}) \sim \mathcal{N}(\hat{\mathbf{K}}^T\mathbf{K}^{-1}\mathcal{F},\hat{\hat{\mathbf{K}}} - \hat{\mathbf{K}}^T\mathbf{K}^{-1}\hat{\mathbf{K}})
\end{aligned}
\label{eqn:predict-condt}
\end{equation}

\subsection{Acquisition function}

In each GPR iteration, a new $h$ must be acquired such that the observed $f(h)$ maximally improves the posterior from the previous iteration.
This requires a \textbf{judicious sampling} strategy that optimizes an improvement metric.
``Expected Improvement'' (EI) is one such popular metric that we will build on in this paper.


\subsubsection{Expected Improvement:}
Given previous observations $\mathcal{D} = (\mathcal{X},\mathcal{F})$, let $f^* = \min_{x \in \mathcal{X}}f(x)$ be the current function minimum (i.e., the minimum observed till now). 
If a new observation $f(h)$ is made at $h$, then the  minimum now will be one of these:
\begin{itemize}
    \item $f(h)$ if $f(h) \leq f^*$
    \item $f^*$ if $f(h) \geq f^*$
\end{itemize}
Hence, the improvement from observing $f$ at $h$ is \\
$[f^*-f(h)]^+$, where, $a^+ = \max(a,0)$.
\new 

We want to choose $h$ that maximizes this improvement. 
However, $f(h)$ is unknown until the observation is made, so we choose $h$ that maximizes the expectation of this improvement. 
\textit{Expected Improvement} is thus defined as:
\begin{equation}
\begin{aligned}
\textbf{EI}(h|\mathcal{X},\mathcal{F}) = \mathbf{E}[[f^*-f(h)]^+|\mathcal{X},\mathcal{F}]
\end{aligned}
\label{eqn:ei}
\end{equation}
where, $\mathbf{E}[\cdot|\mathcal{X},\mathcal{F}]$ is the expectation taken on the GPR posterior distribution given observations $(\mathcal{X},\mathcal{F})$. This posterior is as specified in Eqn. \ref{eqn:posterior}. 
Thus, the next sample to make an observation at is:
\begin{equation}
\begin{aligned}
h = \argmax_{h_i \in \mathbf{R}^N} \textbf{EI}(h_i|\mathcal{X},\mathcal{F})
\end{aligned}
\end{equation}

\subsection{High dimensional Bayesian Optimization}

\textbf{Curse of Dimensionality}: 
The objective function in Eqn. \ref{eqn:opti-prob} typically lies in a high dimensional space (i.e., $h \in \mathcal{H} \subseteq \mathbf{R}^N, N \ge 500$). 
Bayesian optimization works well for functions of $<20$ dimensions \cite{frazier2018tutorial}; with more dimensions, the search space $\mathcal{H}$ increases exponentially, and finding the minimum with {\em few} evaluations becomes untenable. 
One approach to reducing the number of queries is to exploit the sparsity inherent in most real-world functions. 
\new 

We assume our function in Eqn. \ref{eqn:opti-prob} is sparse, i.e., \hl{there is a low-dimensional space that compactly describes $f$}, 
so $f$ has ``low effective dimensions". 
We review ALEBO \cite{letham2020re}, a class of methods that exploit sparsity to create a low-dimensional embedding space using random projections.
Our proposed idea builds on top of ALEBO, but we are actually agnostic of any specific sparsity method.
\subsection{Linear Embedding using Random Projections}
\textbf{Random Projections}: 
Given a function $f: \mathbf{R}^N \to \mathbf{R}$ with effective dimension $d_f$, ALEBO's linear embedding algorithm uses random projections to transform $f$ to a lower dimensional embedding space.
This transformation must guarantee that the minimum $h^*$ from high dimensional space $\mathcal{H}$ gets transformed to its corresponding minimum $y^*$ in low dimensional embedding space. 
The right side of Figure \ref{fig:oracle} aims to visualize this transformation.
Without satisfying this property, optimization in low-dimensions is not possible.
\new


The random embedding is defined by an embedding matrix $\mathbf{B} \in \mathbf{R}^{d \times N}$ that transforms $f$ into its lower dimensional equivalent $f_B(y) = f(h) = f(\mathbf{B}^\dagger y)$, where $\mathbf{B}^\dagger$ is the pseudo-inverse of $\mathbf{B}$. 
Bayesian optimization of $f_B(y)$ is performed in the lower dimensional space $\mathbf{R}^d$. 

\new

\textbf{Clipping to $\mathcal{H}$}: 
When $f$ is optimized over a compact subset $\mathcal{H} \subseteq \mathbf{R}^N$, we cannot evaluate $f$ outside $\mathcal{H}$. 
One approach to prevent any embedding point $y$ from being projected outside of $\mathcal{H}$ is to ``clip" such points to $\mathcal{H}$. 
This is done by projecting the points back into $\mathcal{H}$, i.e., $f_B(y) = f(p_{\mathcal{H}}(\mathbf{B}^\dagger y))$ where, $p_{\mathcal{H}}:\mathbf{R}^N \to \mathbf{R}^N$ is the clipping projection. 
However, this clipping to $\mathcal{H}$ causes nonlinear distortions. 
\new

Instead, constraining the optimization to only points in $\mathcal{Y}$ that do not project outside $\mathcal{H}$, i.e., $\mathbf{B}^\dagger y \in \mathcal{H}$, prevents distortions; 
however, it also reduces the probability of the embedding containing the optimum $h^*$. 
ALEBO remedies this by choosing $d > d_f$ (an embedding space larger than $f$'s effective dimensions). 
Then, the acquisition function evaluated in the constrained embedding space is given as:
    \begin{equation}
    \begin{aligned}
    \argmax_{y \in \mathbf{R}^{d}} \textbf{EI}(y)
    \\
    \textrm{s.t.} \quad -1 \leq \mathbf{B}^\dagger y \leq 1
    \end{aligned}
    \label{eqn:alebo}
    \end{equation}
where the constraint $-1 \leq \mathbf{B}^\dagger y \leq 1$ are linear and form a polytope. 
\new

\textbf{Modifications to the Kernel}: 
ARD kernels in $\mathcal{H}$ (shown in Eqn. \ref{eqn:kernel}) do not translate to a product kernel in embedding $\mathcal{Y}$, since each dimension in $\mathcal{H}$ is independent (diagonal matrix $\Sigma$ in Eqn. \ref{eqn:kernel}).
However, moving along one dimension in embedding is similar to moving across all dimensions of $\mathcal{H}$. To combat this, a {\em Mahalanobis Kernel} is used in the embedding. Any two points in the embedding are projected up to $\mathbf{R}^N$ ($\mathbf{B}^\dagger$) and then projected down to $\mathcal{H}$ ($\mathbf{A}$), $f_B(y)= f(\mathbf{B}^\dagger y) = f(\mathbf{A}\mathbf{B}^\dagger y)$ and $\text{Cov}[f_B(y),f_B(y')] = \exp\{-(y-y')^T\boldsymbol\Gamma(y-y')\}$ where $\boldsymbol\Gamma = (\mathbf{A}^T\mathbf{B}^\dagger)^T\Sigma(\mathbf{A}^T\mathbf{B}^\dagger)$ is a symmetric positive definite matrix.
This finally ensures correctness in sparsity-based Bayesian Optimization (BO).

\new

With this review of Bayesian Optimization and sparsity-based ALEBO, we discuss our algorithm, {\name}.

\section{{\name}}
{\name}'s main contribution is in modifying ALEBO's acquisition function (Eqn. \ref{eqn:alebo}) to incorporate queries of type $Q_d$, namely {\em dimension queries}. 
ALEBO and related sparsity-based algorithms are designed to use filter queries of type $Q_f$; modifying these algorithms to also incorporate $Q_d$ needs cautious design. 
This is because filter samples $h_i$ are $N$-dimensional vectors while dimension queries $h^*[j]$ are scalar values.
Combining these modalities correctly, especially through the sparsity transformations in ALEBO, requires reworking at the heart of sparsity-based BO methods.


\new

\begin{figure*}[htbp]
{\includegraphics[width=0.70\textwidth]{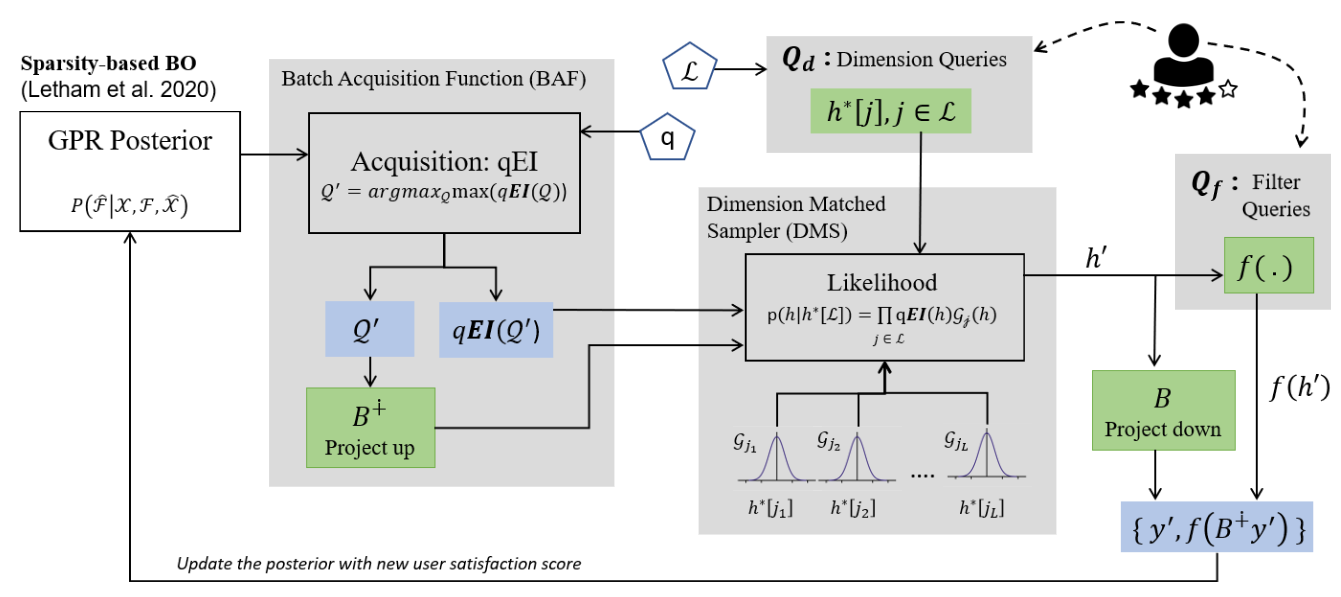}}
\hfill
{\includegraphics[width=0.28\textwidth]{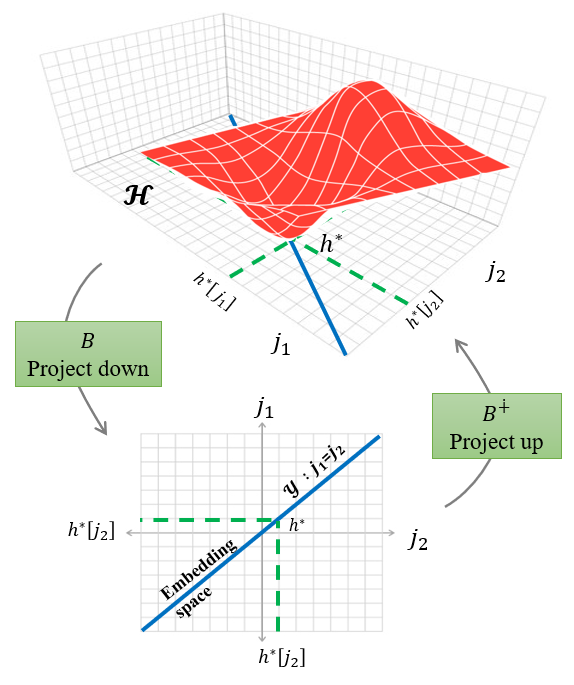}}
\caption{{System flow:  {\name} consists of three modules: \textbf{BAF}, \textbf{DMS}, and GPR posterior. Green boxes denote the system inputs and hyper-parameters, and blue marks the module outputs.
The right figure shows the transformation between the high and low dimensional spaces, made feasible by the random embedding matrix in ALEBO.}}
\label{fig:oracle}
\end{figure*}

Figure \ref{fig:oracle}(a) illustrates the design of {\name} -- the modules in gray are the proposed extensions over the literature.
The two key modules are 
(1) \textbf{Batch Acquisition Function (BAF)}, and 
(2) \textbf{Dimension Matched Sampler (DMS)}. 
\new 

Conventional SparseBO obtains a {\em single} sample $h'$ from the acquisition function and makes a function observation $f(h')$. 
This observation is then used to update the GPR posterior. 
In contrast, the \textbf{BAF} module in {\name} intends to pick a batch of $q$ (jointly optimal) samples $\{h'_1,h'_2,\dots,h'_q\}$ and the \textbf{DMS} module orders them preferentially by matching them against the dimensional information of the minimizer $h^*[j], j \in \mathcal{L}$. 
Through this method, we are selecting the next filter sample $h'$ by essentially combining the ``wisdom" of both types of querying. 
Said differently, we first sample a batch of candidates which are all ``good" choices as per the EI metric \hl{(e.g., each sample is a different sushi recipe)}, and then, dimension matching makes the final selection in favor of one $h'\in \{h'_1,h'_2,\dots,h'_q\}$ that aligns with the optimal $h^*[j], j \in \mathcal{L}$.
This means if the user has a high preference for the sweet dimension, then the next $Q_f$ query $h'$ becomes a ``sweet sushi recipe".
The details are presented next.

\subsection{Batch Acquisition Function (BAF)}
Instead of picking one sample $h'$, \textbf{BAF} proposes to pick a batch of $q$ samples $\{h'_1,h'_2,\dots,h'_q\} = \mathbf{B}^\dagger \{y'_1,y'_2,\dots,y'_q\}$ that {\em jointly} maximize the acquisition function.

\new

We assign a joint metric, {\em q-ExpectedImprovement} (qEI), to a set of $q$ candidate points $\mathcal{Q}'=\{y'_1,y'_2,\dots,y'_q\}$ in the search space $\mathcal{Y}$. 
This is realized through two steps: 
\begin{itemize}
    \item[(1)] Sampling the joint distribution of $q$ points under the current posterior using MCMC sampling \cite{neal2003slice} to obtain $\mathcal{Q}=\{y_1,y_2,\dots,y_q\}$
    \item[(2)] Evaluating joint metrics $\text{q}\textbf{EI}(y_i)$ for $y_{1:q}$ over the current minimum $f^* = \text{min}_{y \in \mathcal{Y}}f(\mathbf{B}^\dagger y)$ as follows:
\end{itemize} 
\begin{equation}
\begin{aligned}
\text{q}\textbf{EI}(y_i|\mathcal{D}_n,\mathcal{Q}) &= \mathbf{E}[[f^*-f(y_i)]^+|\mathcal{D}_n,\mathcal{Q}]\\
\text{q}\textbf{EI}(\mathcal{Q}) &\triangleq \{\text{q}\textbf{EI}(y_i|\mathcal{D}_n,\mathcal{Q})\}
\end{aligned}
\label{eqn:qei}
\end{equation}
    
where $\mathcal{D}_n$ denotes current set of observations. 
Compared to $\textbf{EI}(y_i|\mathcal{D}_n)$ in Eqn.~\ref{eqn:ei},  $\text{q}\textbf{EI}(y_i)$ marginalizes the expectation over the candidate set $\mathcal{Q}$.
\new 

We select the candidate set $\mathcal{Q}'$ of highest expected improvement as follows:
\begin{equation} 
\begin{aligned}
\mathcal{Q}'=\{y'_1,y'_2,\dots,y'_q\} &= \argmax_{\mathcal{Q} = \{y_1,y_2,\dots,y_q\}}  max(\text{q}\textbf{EI}(\mathcal{Q}))\\
\text{s.t.\ } -1 &\leq \mathbf{B}^\dagger y_i' \leq 1\ \ \ \forall y_i'\in\mathcal{Q}'
\end{aligned}
\label{eqn:qei1}
\end{equation}
where each $\mathcal{Q}$ is ranked with the highest $\text{q}\textbf{EI}(y_i)$ for $y_i\in\mathcal{Q}$.
\hl{Note that $\mathcal{Q}$ is subject to the constraint} from Eqn. \ref{eqn:alebo}; this ensures \textbf{BAF} operates within the bounds of $\mathcal{H}$ under ALEBO's random projections to lower-dimensional space, $\mathbf{R}^d$.
\new 

These points and their corresponding q\textbf{EI} metric values $(\mathcal{Q}',\text{q}\textbf{EI}(\mathcal{Q}'))$ are then passed as input to \textbf{DMS}. 

\subsection{Dimension Matched Sampler (DMS)}
\textbf{DMS}'s task is to output one sample $h' = \mathbf{B}^\dagger y'$ that best matches the dimensional information of the minimizer $h^*[j], j \in \mathcal{L}$ obtained from $Q_d$ type queries. This $h'$ will be the sample at which the $Q_f$ filter query is made to update the GPR posterior. 
Clearly, the \textbf{DMS} algorithm must operate in high-dimensional space $\mathcal{H}$ as both $h'$ and $h^* \in \mathcal{H}$. 
In contrast, the \textbf{BAF} module operates in the $d-$dimensional embedding space $\mathbf{R}^d$, hence the $q$ candidates lie in this embedding space. 
To remedy this, the \textbf{BAF}'s outputs in the embedding space are projected up to $\mathcal{H}$, i.e., $\mathcal{Q}'=\{y'_1,y'_2,\dots,y'_q\} \to \mathcal{T}'=\{h'_1,h'_2,\dots,h'_q\}$  as shown in Figure \ref{fig:oracle}(a) (the green box labeled $\mathbf{B}^\dagger$). Figure~\ref{fig:oracle}(b) illustrates the translation of any point from high-dimensional space $\mathcal{H}$ to the low-dimensional embedding space $\mathcal{Y}$ and vice versa. 
\hl{Since \textbf{DMS} utilizes \textbf{BAF}'s} q\textbf{EI} metric, the $q$ chosen candidate samples $\mathcal{Q}'$ adhere to the box constraints in Eqn. \ref{eqn:alebo}. 
This ensures that the $q$ candidates in high-dimensional space $\mathcal{T}'$ lie inside $\mathcal{H}$, thereby avoiding any non-linearity due to clipping.

\new

Once $\mathcal{Q}'$ has been translated to higher dimensional $\mathcal{T}'$, \textbf{DMS} uses a joint {\em likelihood} measure to preferentially order the $q$ samples based on their degree of similarity to the $L$ dimension queries $h^*[j], j \in \mathcal{L}$ as shown in Eqn. \ref{eqn:DMS}. Given $h^*[j], j \in \mathcal{L}$ queries, \textbf{DMS}' joint likelihood measure computes the likelihood of a filter sample $h$ maximally improving the GPR posterior when observed.

\begin{equation}
\begin{aligned}
P(h|h^*[\mathcal{L}]) &= \prod_{j \in \mathcal{L}} \text{q}\textbf{EI}(h) \mathcal{G}_j(h)\\
h' &= \argmax_{h \in \mathcal{T}'} P(h|h^*[\mathcal{L}])
\end{aligned}
\label{eqn:DMS}
\end{equation}
where, $\mathcal{G}_j = \mathcal{N}(\mu=h^*[j],\sigma)$ is a Gaussian distribution with mean $h^*[j]$ and variance $\sigma$ for each $Q_d$ dimension available $j\in \mathcal{L}$, and $\text{q}\textbf{EI}(h'_i)$ is the \textbf{BAF} acquisition metric of a candidate sample $h'_i$. 
\new

The maximizer in Eqn. \ref{eqn:DMS}, $h'$, is the sample at which the user is queried.
This $h'$ and the user's satisfaction score $f(h')$ are then projected down to the embedding as $(y',f(\mathbf{B}^\dagger y'))$.
Finally, this tuple is used to update the GPR posterior for the next iteration of {\name}.

\section{Experiment: Synthetic BlackBox Functions}
\label{sec:Obj}
We first present experiments on various synthetic functions $f(h)$.
We pretend the function is a black-box, but filter and dimension queries are feasible.
Obviously, because we actually know the function, we will evaluate how close {\name} can get to the global minima.
\new 

\begin{figure}[htbp]
\centerline{\includegraphics[width=0.35\textwidth]{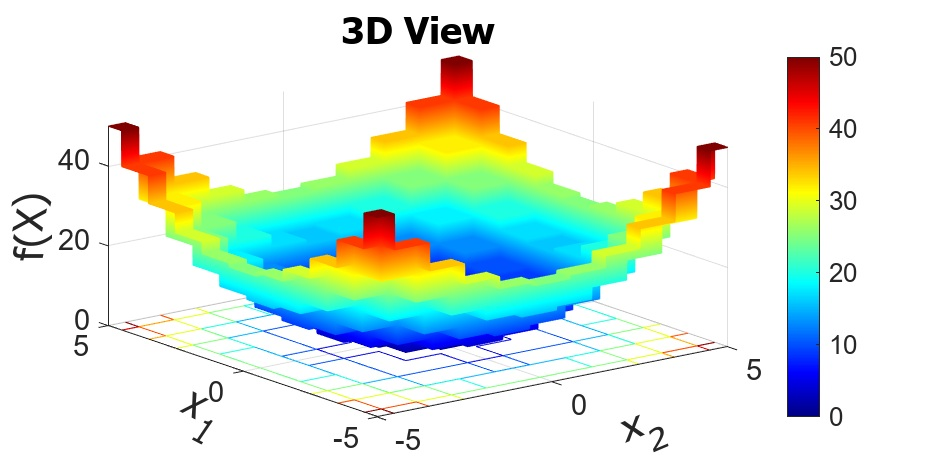}}
\caption{Satisfaction Function $P1$: Discontinuous staircase structure, containing infinite zero gradient regions.}
\label{fig:P1}
\end{figure}

We evaluate two sets of objective functions: \\
\textbf{(1) Staircase Satisfaction Functions} that are shaped like a staircase (see Figure \ref{fig:P1}) and roughly mimic how humans rate their experiences in discrete steps \cite{Al-Roomi2015}. \\
\textbf{(2) Benchmark Functions} commonly used in Bayesian optimization research~\cite{ssurjano}, such as \texttt{BRANIN}, \texttt{HARTMANN6}, and \texttt{ROSENBROCK}. 

\subsubsection{Baseline and Metrics:}
We consider a baseline that extends ALEBO with the additional information from $L$ dimensional queries.
This implies that ALEBO's search space can be reduced from $\mathcal{R}^N$ to $\mathcal{R}^{N-L}$.
Our evaluation metric is {\em Regret}, which is the difference between the predicted minimum and true global minimum $({f}(\hat{h}^*) - f(h^*))$. 
All reported results are an average of $10$ different runs. 
More details on the objective functions and evaluation parameters are included in the Appendix.
\new 

In the following figures, X-axis label ``function evaluations'' indicates the number of filter queries ($Q_f$), each of which produces a GPR iteration.
Also, $L$ denotes the number of dimension queries ($Q_d$).
For comparison, we mark points on the graph that use the same query budget, $B=Q_f+Q_d$\footnote{For readability, we abuse the notation $Q_d$, which is equal to the number of dimension queries, $L$.}.
\new 

{\textbf{Comparison to ALEBO($L$)}:
Figure \ref{fig:naive} shows the performance of {\name}  against ALEBO($L$).
ALEBO($L=0$) shows the weakest performance because it does not benefit from dimensional queries.
With $5$ dimensional queries, ALEBO($L=5$) shows immediate gain since it has to only search $\mathbf{R}^{N-5}$.
{\name} shows further improvement with $L=5$, implying that the combination of $Q_f$ and $Q_d$ queries are beneficial, even though the search space is $\mathcal{R}^N$.
Observe that points marked with stars all have the same query budget $B=90$, thus, {\name} achieves high satisfaction (low regret) for a given $B$.
Of course, if $B$ is too small, say $<30$, then the gains reduce.
This is understandable because the GPR posterior has not yet converged. 
Finally, when $L$=$15$, the regret is even lower. 
\new 
\begin{figure}[htbp]
\includegraphics[width=0.9\columnwidth]{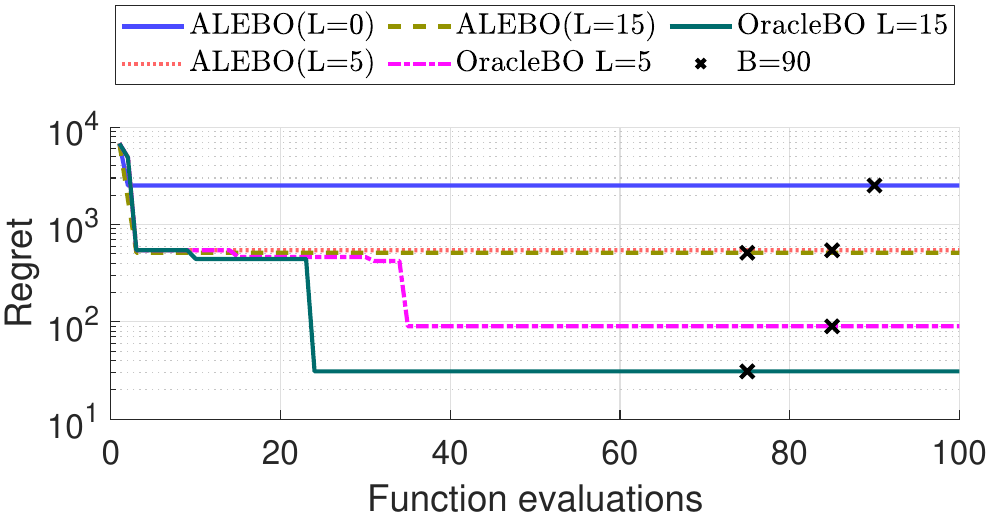}
\caption{Performance on ALEBO($L$) and \name.}
\label{fig:naive} 
\end{figure}
\begin{figure}[htbp]
\includegraphics[width=0.9\columnwidth]{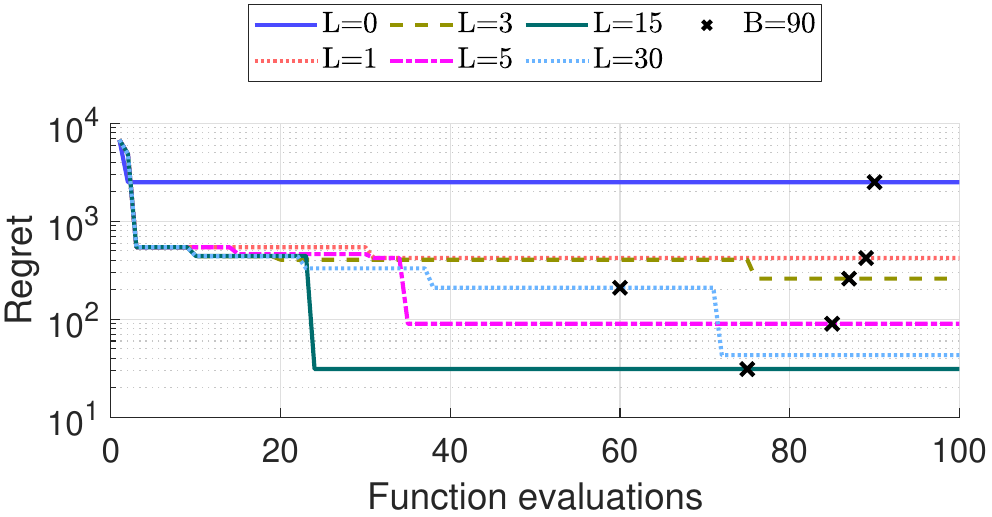}
\caption{Different number of $Q_d$ queries on \name.}
\label{fig:q1cmp}
\end{figure}

\textbf{Effect of Varying $L$}: 
Figure \ref{fig:q1cmp} reports the impact of increasing dimension queries, $L$, on regret. 
Observe that for a fixed query budget $B=90$, increasing $L$ is beneficial but only up to $L = 15$.
Increasing $L$ further offers more information about the optimal $h^*$ but at the expense of lowering the number of $Q_f$ queries.
Evidently, for the staircase function, the empirical optimal for $L$ is in the neighborhood of $15$.
\new

\textbf{Which L out of N queries?}
Given $L=15$ queries, say, different subsets of $N$ dimensions can be chosen. 
Let us denote this subset as $\mathcal{L}$.
If $f(h)$ hardly varies along the dimensions included in $\mathcal{L}$, then $\mathcal{L}$ contributes little to estimating the satisfaction function.
Figure ~\ref{fig:importance} shows \name's regret on two different $\mathcal{L}$. 
Note that because the objective function $f$ is synthesized, the variation of $f$ against any dimension $y$ is known.
In $\mathcal{L}_{Top}$, we select the $L$ dimensions of 
largest variances; $\mathcal{L}_{Rand}$ denotes the 
randomly selected dimensions from $\{1,N\}$.
Results show that $\mathcal{L}_{Top}$ achieves lower regret (median and variance) compared to $\mathcal{L}_{Rand}$.
Thus, in real applications, it helps to choose $L$  dimensions that are likely to influence the user's satisfaction

\begin{figure}[htbp]
\includegraphics[width=\columnwidth]{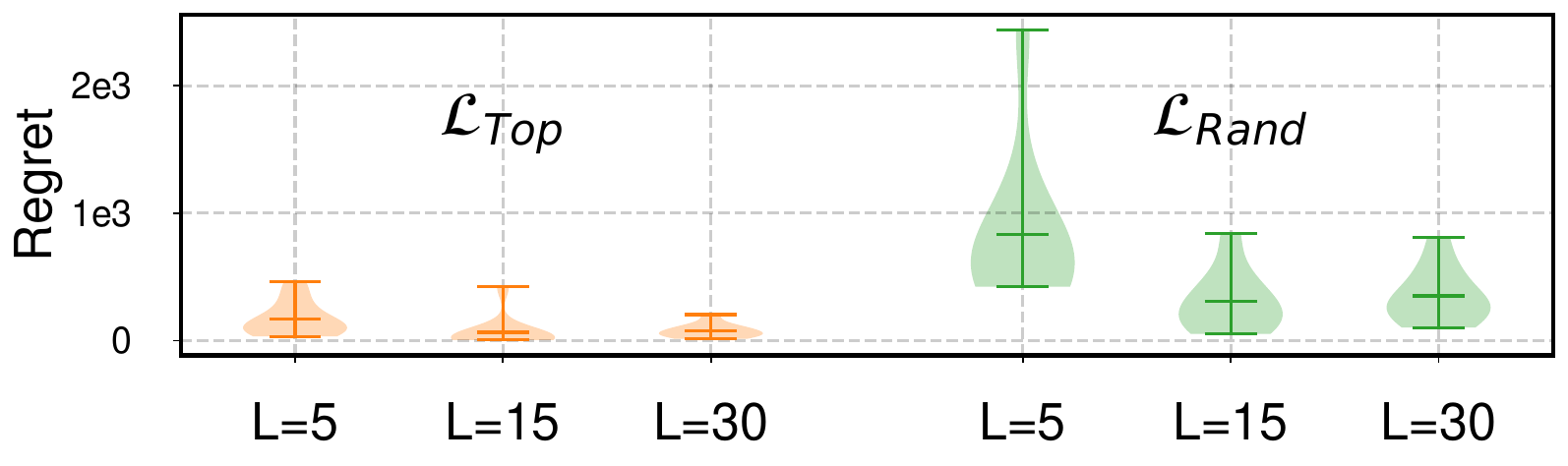}
\caption{Distribution of {\name}'s regret on different $Q_d$ subsets $\mathcal{L}_{Top}$ and $\mathcal{L}_{Rand}$.}
\label{fig:importance}
\end{figure}



\hl{\textbf{Hyperparameter Selection}}: 
Parameters in \textbf{BAF} and \textbf{DMS} modules include: $N$, dimension of the filter; $d$, dimension of the embedding; $q$, number of candidates BAF outputs, and $\sigma$, the dimensional variance in \textbf{DMS}.
\new

\begin{table}

\centering
\begin{tabular}{|c|c|c|c|c|c|c|}
\hline
\multirow{2}{*}{N} & \multirow{2}{*}{d} & \multicolumn{5}{c|}{$(q,\sigma)$} \\ 
\cline{3-7} 
& & \multicolumn{1}{c|}{(5,1)} & 
\multicolumn{1}{c|}{(2,0.2)} & 
\multicolumn{1}{c|}{(2,10)} &  
\multicolumn{1}{c|}{(7,0.2)} &  
\multicolumn{1}{c|}{(7,10)} \\ \hline

\multirow{3}{*}{500} & $4$ 
& \textbf{83} & 445 & 316 & 1459 & 1459 \\\cline{2-7} 
 
& $10$ & 148 & 166 & 237 & 760 & 883 \\\cline{2-7}
 
& $20$ & 477 & 551 & 609 & 1201 & 1255\\ \hline

\multirow{3}{*}{2000} & $4$ & \textbf{90} & 242 & 514 & 543 & 628 \\ \cline{2-7}
 
& $10$ & 2166 & 2331 & 2753 & 2331 & 2753\\ \cline{2-7}
 
& $20$ & 2677 & 2764 & 3125 & 2764 & 3125\\\hline

\multirow{3}{*}{4000} & $4$ & 513 &606 & 1268 & 1268 & 1268 \\ \cline{2-7} 
 
& $10$ & 1532 & 1627 & 2154 & 4006 & 5278\\ \cline{2-7}
 
& $20$ & 1749 & 1993 & 3332 & 10213 & 10213\\\hline

\end{tabular}
\caption{Hyperparameter analysis on regret for $P1$.}
\label{tbl:hyper}
\end{table}

\begin{figure*}[htbp]
\includegraphics[width=0.5\textwidth]{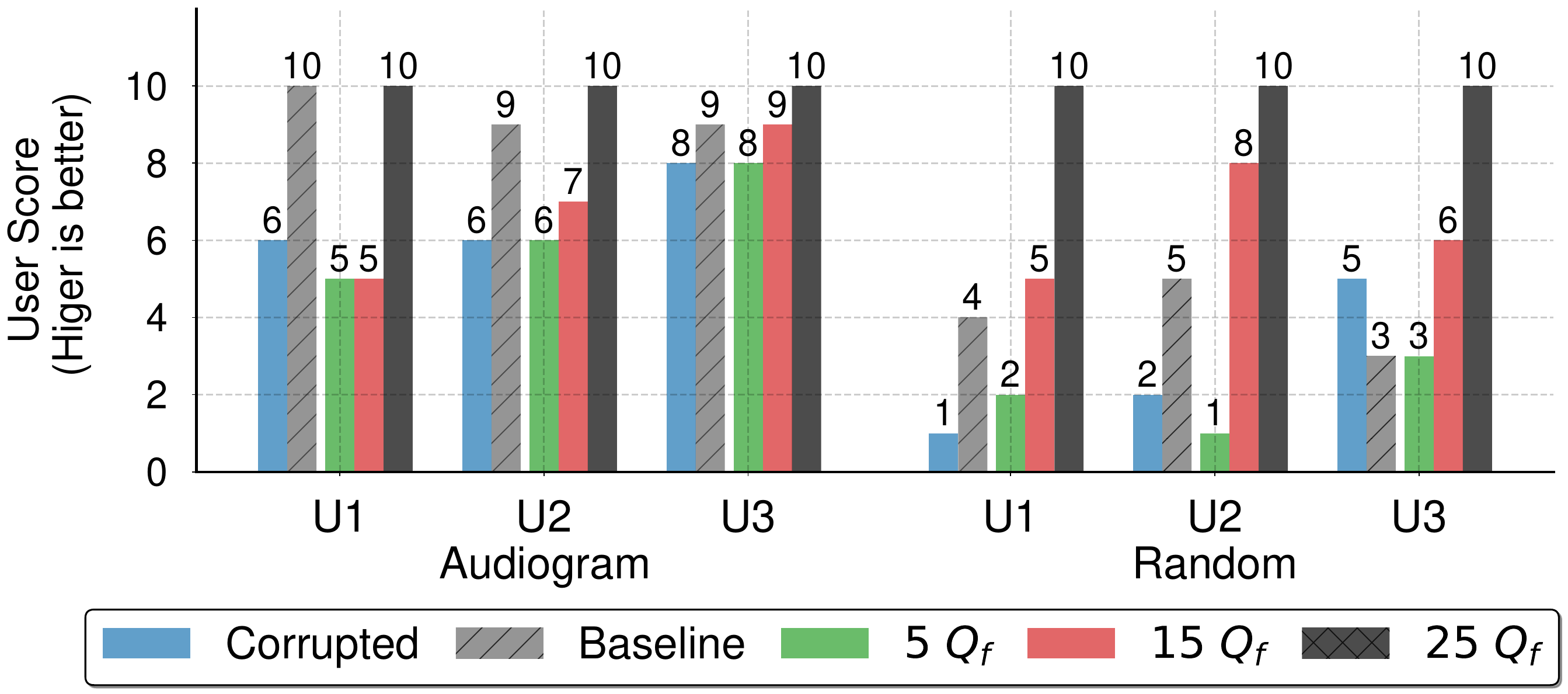}
\hfill
\includegraphics[width=0.5\textwidth]{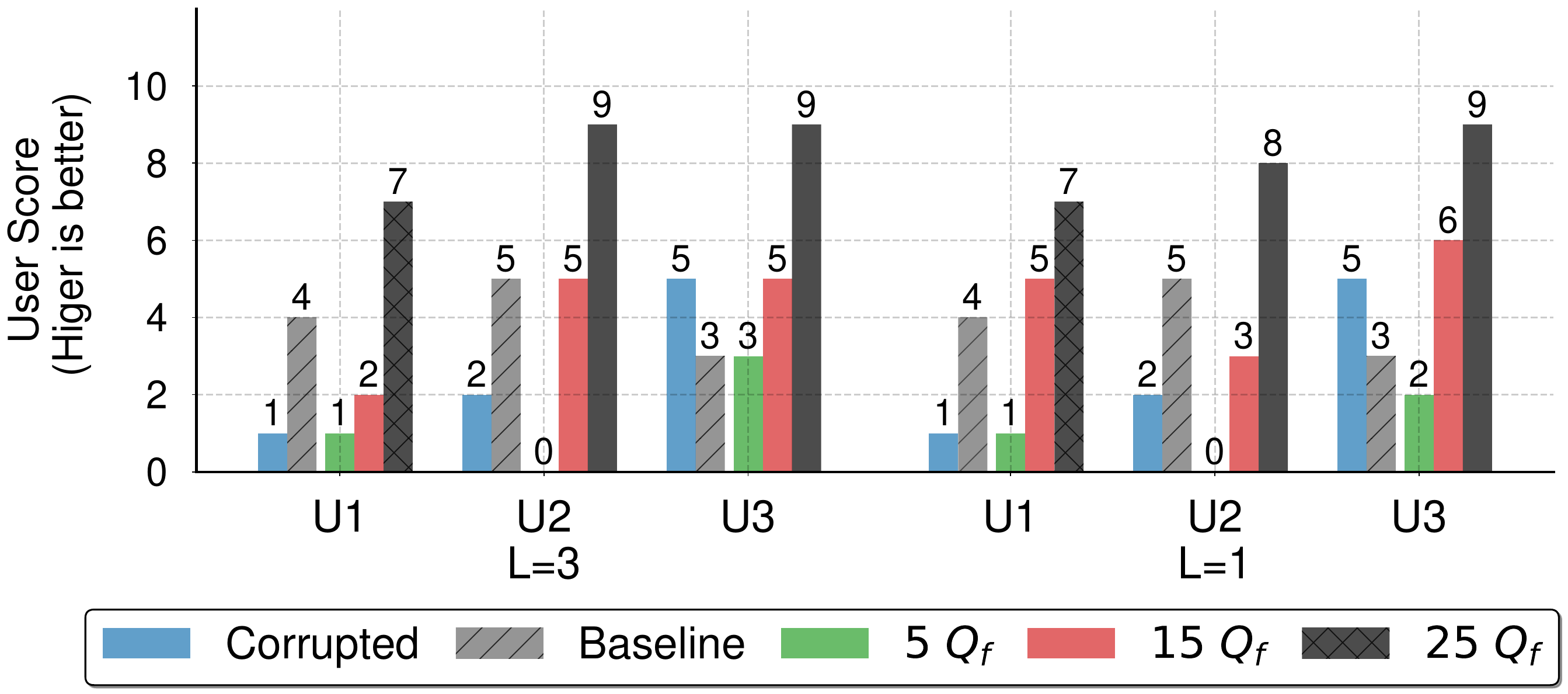}
\caption{User score comparison on (a) $L=5$ on hearing-loss profile and random profile. (b) $L=[1,3]$ on random profile.}
\label{fig:ado_personal}
\end{figure*}

\section{Experiments: Audio Personalization}
This section reports experiments with real volunteers \hl{in the context of personalizing hearing aids}. Today's hearing aids aim to filter the audio with $h$ so that the user's hearing loss is compensated, and their satisfaction $f(h)$ is maximized. 
Hearing aids prescriptions exactly perform the process of dimension querying where different frequency tones $j$ are played to the user, and their optimal audibility is recorded as $h^*[j]$. 
To minimize user burden, audio clinics play around $L$=$7$ frequency tones (from different octaves) and interpolate through them to generate the user's personalized filter. 
This filter is called the {\em audiogram}.
\new

Interpolation is obviously a coarse approximation of the user's true personal filter, $h^*$. 
We expect to improve the user's satisfaction over their {\em audiogram}, using a modest number of $Q_f$ queries prescribed by {\name}. 
In other words, the user can attain higher satisfaction $\hat{f}^*$ if they are willing to listen and rate some audio clips ($Q_f$) at home. 
\new 

For experimentation, we invited $3$ volunteers with no hearing loss. To emulate hearing loss, we played audio that was deliberately ``\texttt{corrupted}'' with hearing loss profiles from the public hearing-loss dataset in NHANES ~\cite{salmon2022audiogram}. 
This corrupted audio obviously yields a poor satisfaction score from our volunteers. 
We compute the \hl{coarse-grained audiogram for the volunteers} and compare {\name} against this ``\texttt{baseline}''. 
\new

\textbf{Results}: 
Figure \ref{fig:ado_personal}(a) plots the final satisfaction score from the $3$ volunteers (U1, U2, U3), first for the ``Audiogram'' experiment, and then for a ``Random'' filter experiment (to be described soon).
The \texttt{Corrupted} signal obviously receives a low score, but the interpolated audiogram, labeled \texttt{Baseline}, considerably improves the score.
{\name} is still able to match/improve user satisfaction with $Q_d=5$ and $Q_f=25$ queries.
With fewer $Q_f$ of $5$ and $15$, {\name} could not outperform \texttt{Baseline} as the function $f$'s search space $\mathcal{R}^{4000}$ had not been sufficiently sampled. 
\new 


The audiogram \texttt{Baseline} performs quite well primarily because human hearing loss is reasonably flat within octaves, hence, interpolation is adequate. 
We thus explore another application that injects more complex audio distortions, e.g., a cheap music speaker.
\new

We again emulate this distortion by deliberately corrupting the audio with a random filter $h$, each $h[j]$ selected independently from [$-30, 30$]dB. 
Similar interpolation as an audiogram, using $L=5$, will give us a new \texttt{Baseline}. 
Fig \ref{fig:ado_personal}(a)-Random plots the results for the same $3$ users.
{\name} improves the satisfaction scores even with $Q_f=15$ queries, and achieves the maximum with $Q_f=20$ queries.
\new

We also investigate the degradation of user satisfaction with fewer dimensional queries $L = 1, 3$. 
Figure \ref{fig:ado_personal}(b) reports the results for only the Random distortion filter.
Evidently, the degradation is graceful, i.e., as $L$ reduces, more $Q_f$ filter queries are needed to achieve the same level of personal satisfaction.
The audio demos at various stages of the optimization are made available\footnote{https://oraclebo.github.io/}.

\section{Related Work}
\label{sec:related-work}
To the best of our knowledge, {\name} is the first work that combines two different types of queries, $Q_f$ and $Q_d$, for Bayesian Optimization. 
We also believe such hybrid querying has not been applied in audio personalization. 
The closest work in the application context is \cite{solnik2017bayesian} where authors used conventional BO to search for the best-rated cookie recipe at Google.
We believe more applications can benefit, and the burden of querying users can become practical with dimension querying ($Q_d$).
Of course, BO has been extensively used to solve complex problems that do not have humans in the loop.
These include material science \cite{ueno2016combo}, medicine \cite{negoescu2011knowledge}, hyperparameter tuning in neural networks \cite{snoek2012practical}, etc.
\new

{\name} inherits sparsity-based BO frameworks based on low-dimensional embeddings. 
These papers use linear random projections to map from high to low-dimensional spaces \cite{qian2016derivative}\cite{wang2016bayesian}\cite{binois2020choice}\cite{letham2020re}. 
Authors of \cite{garnett2013active}\cite{lu2018structured} use a Gaussian Process to simultaneously learn the model and the embedding. 
Non-linear embeddings are learnt using Variation Autoencoders in \cite{gomez2018automatic}\cite{moriconi2020high}\cite{lu2018structured}. 
{\name} is agnostic to these algorithms and we expect their advantages to reflect in our performance as well.
On a similar note, various papers modify the kernel \cite{kandasamy2015high}\cite{gardner2017discovering}\cite{mutny2018efficient}\cite{wang2018batched} to restrict the candidate function choices \cite{oh2018bock}, reflecting additional structure in the objective function.
LineBO \cite{kirschner2019adaptive} optimizes the acquisition function along one-dimensional lines. 
TuRBO \cite{eriksson2019scalable} employs trust regions around the current minimizer. \cite{oh2019combinatorial}, \cite{eriksson2021high} use sparsity creating prior. 
The performance of {\name} can be boosted with such kernel manipulations.

\section{Follow-up Work and Conclusion}
This paper adds a new type of querying to black-box optimization where an Oracle can reveal information about the minimizer, $h^*$. 
This type of dimension querying maps to practical applications and solutions that achieve a low query budget $B$ can be useful.
This paper is a first step in this direction, starting with an empirical treatment of the problem.
We believe the findings are promising and open doors to follow-up work.
For instance, (1) an analytical treatment on convergence is needed for the hybrid $Q_f + Q_d$ querying. 
(2) If $j$ can be freely chosen in $h^*[j]$, how many and which $j$'s are optimal, given the query budget $B$? 
(3) What other applications lend themselves to the notion of hybrid querying? 
(4) Dimension queries may not independent, i.e., a user may like sugar for a certain recipe and may not like sugar for another recipe. How can such conditionals be incorporated in the formulation?
We hope {\name} serves as a stepping stone to solving important problems along these directions.

%% file: appendix.tex
\newpage
\appendix
\section{Technical Appendix}
\label{sec:append}

\subsection{Synthetic Functions Experiment Details}
In this section, we discuss the experiment setup for {\name}'s application to synthetic satisfaction functions and other BO benchmark functions. 
Since we have knowledge of the minimizer $h^*$, we simulate dimension querying $Q_d$ at queried dimensions $\mathcal{L}$ i.e., $h^*[j], j \in \mathcal{L}$.

The two sets of objective functions to be optimized are:
\begin{itemize}
    \item[(1)]  \textbf{Satisfaction Functions}: We generated functions \cite{Al-Roomi2015} that have a discontinuous staircase structure to mimic the user satisfaction scoring function. The staircase structure is due to the fact that a user's audio perception might not change for a range of filters so the score remains the same and might change with sudden jumps for some filter choices, thus leading to a flat shape in some regions and steep curve in other regions. The staircase structure results in the functions having infinite local minima, infinite global minima, and zero gradient regions. These functions are naturally not suited for gradient-based optimization techniques. In our work, we use three such perception functions denoted as $P1, P2$, and $P3$. The functions are defined in Eqns \ref{eqn:P1},\ref{eqn:P2},\ref{eqn:P3}.
    \item[(2)] \textbf{Benchmark Functions}: We test {\name} on commonly used benchmark functions in Bayesian optimization research: BRANIN, HARTMANN6, and ROSENBROCK \cite{ssurjano}. These functions are denoted as $B, H$, and $R$. The functions are defined in Eqns \ref{eqn:B},\ref{eqn:H},\ref{eqn:R}.
\end{itemize}

\begin{equation}
\begin{aligned}
f_{P1}(\mathbf{h}) = \sum_i^N (\lfloor |h_i + 0.5| \rfloor)^2
\end{aligned}
\label{eqn:P1}
\end{equation}
where, $-100\le h_i \le 100, i=1,2,\dots,N$, $h_i$ is filter $h$ along dimension $i$. Infinite global minima at $f_{min}(\mathbf{h}^*) = 0$, and the minimizers are $-0.5 \le h_i^* < 0.5$ (i.e.,) $h_i^* \in [-0.5,0.5), i=1,2,\dots,N$
\begin{equation}
\begin{aligned}
f_{P2}(\mathbf{h}) = \sum_i^N (\lfloor |h_i| \rfloor)
\end{aligned}
\label{eqn:P2}
\end{equation}
where, $-100\le h_i \le 100, i=1,2,\dots,N$. Infinite global minima at $f_{min}(\mathbf{h}^*) = 0$, and the minimizers are $-1 < h_i^* < 1$ (i.e.,) $h_i^* \in (-1,1), i=1,2,\dots,N$
\begin{equation}
\begin{aligned}
f_{P3}(\mathbf{h}) = \sum_i^N (\lfloor (h_i)^2 \rfloor)
\end{aligned}
\label{eqn:P3}
\end{equation}
where, $-100\le h_i \le 100, i=1,2,\dots,N$. Infinite global minima at $f_{min}(\mathbf{h}^*) = 0$, and the minimizers are $-1 < h_i^* < 1$ (i.e.,) $h_i^* \in (-1,1), i=1,2,\dots,N$

\begin{equation}
\begin{aligned}
f_{B}(\mathbf{h}) = a(h_2 - bx_1^2 +cx_1 -r)^2 + s(1-t)cos(x_1) + s
\end{aligned}
\label{eqn:B}
\end{equation}
where, $h_i\in[-5,10], h_2\in[0,15], a=1, b=5.1⁄(4\pi^2), c=5⁄\pi, r=6, s=10, t=1⁄(8\pi).$. Three global minima at $f_{min}(\mathbf{h}^*) = 0.397887$, and the minimizers are $\mathbf{h}^*=(-\pi,12.275), (\pi,2.275), (9.42478,2.475)$
\begin{equation}
\begin{aligned}
f_{H}(\mathbf{h}) = \sum_i^4 \alpha_i \exp{-\sum_j^6(A_{ij}(h_j - P_{ij})^2}
\end{aligned}
\label{eqn:H}
\end{equation}
where, $h_i\in(0,1), i=1,2\dots,6$, $\alpha=(1,1.2,3,3.2)^T$, $A=\begin{bmatrix}
    10 & 3 & 17 & 3.5 & 1.7 & 8 \\0.05 & 10 & 17 & 0.1 & 8 & 14 \\3 & 3.5 & 1.7 & 10 & 17 & 8 \\17 & 8 & 0.05 & 10 & 0.1 & 14 \\
\end{bmatrix}$, $P=\begin{bmatrix}
    1312 & 1696 & 5569 & 124 & 8283 & 5886 \\2329 & 4135 & 8307 & 3736 & 1004 & 9991 \\2348 & 1451 & 3522 & 2883 & 3047 & 6650 \\4047 & 8828 & 8732 & 5743 & 1091 & 381 \\
\end{bmatrix}$. One global minimum at $f_{min}(\mathbf{h}^*)=-3.32237$, and the minimizer is $\mathbf{h}^*=(0.20169,0.150011,0.476874,0.275332,0.311652,0.6573)$.
\begin{equation}
\begin{aligned}
f_{R}(\mathbf{h}) = \sum_i^{N-1} (100(h_{i+1} - h_i^2)^2 + (h_i -1)^2)
\end{aligned}
\label{eqn:R}
\end{equation}
where, $h_i\in[-5,10], i=1,2\dots,N$. One global minimum at $f_{min}(\mathbf{h}^*)=0$, and the minimizer is $h_i^*=1, i=1,2\dots,N$.

\subsubsection{Baseline and Metrics:}
We consider a baseline that extends ALEBO with the additional information from $L$ dimensional queries.
This implies that ALEBO's search space can be reduced from $\mathcal{R}^N$ to $\mathcal{R}^{N-L}$.
Our evaluation metric is {\em Regret}, which is the difference between the predicted minimum and true global minimum $({f}(\hat{h}^*) - f(h^*))$. 

\subsubsection{Evaluation Parameters}
\label{sec:app_para}
In our experiments, we optimize the functions for:
\begin{itemize}
    \item[-] f\_evals = 100 function evaluations
    \item[-] r\_init = 5 initial random samples after which the acquisition sampling begins
    \item[-] $N = 2000, \mathcal{H} \subseteq \mathbf{R}^N$ is the high-dimensional space
    \item[-] $\mathbf{R}^d, d = 4$ is embedding space for $P1, P2, P3$, Branin and Rosenbrock and is $\mathbf{R}^d, d = 6$ for Hartmann6 . Branin, Rosenbrock, and Hartmann6 have effective dimensionality $d_e = 2, 4, 6$, respectively. 
    \item[-] Different numbers of $Q_d$ queries $L = {0, 1, 3, 5, 15, 30}$ are used and $L=0$ implies no $Q_d$ queries are available and {\name} functions as just ALEBO
    \item[-] $\mathcal{L}$ $Q_d$ Top and Random dimensions for $Q_d$ queries are considered.
    \item[-] $q=5$ acquisition samples are used in Batch Acquisition Function in Eqns \ref{eqn:qei},\ref{eqn:qei1}
    \item[-] In the Dimension Matched Sampler, we use variance $\sigma = 1$: $\mathcal{G}_j = \mathcal{N}(\mu=h^*[j],\sigma=1)$ for each dimension $j\in \mathcal{L}$ in Eqn \ref{eqn:DMS}
    \item[-] For Branin, we use the minimizer $\mathbf{h}^*=(\pi,2.275)$ to generate dimension $Q_d$ queries and for perception functions $P1,P2,P3$ we use the minimizer $h_i^*=0, i=1,2,\dots,N$. For Hartmann6 and Rosenbrock we use their unique minimizers 
    \item[-] We run 10 random runs of each experiment
    
\end{itemize} 
\new 

\begin{equation}
\begin{aligned}
\text{q}\textbf{EI}(y_i|\mathcal{D}_n,\mathcal{Q}) &= \mathbf{E}[[f^*-f(y_i)]^+|\mathcal{D}_n,\mathcal{Q}]\\
\text{q}\textbf{EI}(\mathcal{Q}) &\triangleq \{\text{q}\textbf{EI}(y_i|\mathcal{D}_n,\mathcal{Q})\}
\end{aligned}
\label{eqn:qei}
\end{equation}
where $\mathcal{D}_n$ denotes current set of observations. 
$\text{q}\textbf{EI}(y_i)$ marginalizes the expectation over the candidate set $\mathcal{Q}$.
\new 

We select the candidate set $\mathcal{Q}'$ of highest expected improvement as follows:
\begin{equation} 
\begin{aligned}
\mathcal{Q}'=\{y'_1,y'_2,\dots,y'_q\} &= \argmax_{\mathcal{Q} = \{y_1,y_2,\dots,y_q\}}  max(\text{q}\textbf{EI}(\mathcal{Q}))\\
\text{s.t.\ } -1 &\leq \mathbf{B}^\dagger y_i' \leq 1\ \ \ \forall y_i'\in\mathcal{Q}'
\end{aligned}
\label{eqn:qei1}
\end{equation}
where each $\mathcal{Q}$ is ranked with the highest $\text{q}\textbf{EI}(y_i)$ for $y_i\in\mathcal{Q}$.
\begin{equation}
\begin{aligned}
P(h|h^*[\mathcal{L}]) &= \prod_{j \in \mathcal{L}} \text{q}\textbf{EI}(h) \mathcal{G}_j(h)\\
h' &= \argmax_{h \in \mathcal{T}'} P(h|h^*[\mathcal{L}])
\end{aligned}
\label{eqn:DMS}
\end{equation}
where, $\mathcal{G}_j = \mathcal{N}(\mu=h^*[j],\sigma)$ is a Gaussian distribution with mean $h^*[j]$ and variance $\sigma$ for each $Q_d$ dimension available $j\in \mathcal{L}$, and $\text{q}\textbf{EI}(h'_i)$ is the \textbf{BAF} acquisition metric of a candidate sample $h'_i$. 

\subsection{Audio Personalization Experiment Details}
\label{sec:app_aud}
\ul{\textbf{Queries}}: The detailed definitions of queries on audio personalization are as follows:
\begin{itemize}
\item[1.] \textbf{Query1 ($Q_d$): A clinical audiogram test}: Conventional hearing aids tuning involves playing pure-tone frequencies to the patient and capturing their hearing response. For each frequency $j$ played with increasing amplitude (-10 dB to 120 dB), the user presses a button when they can hear the frequency. Doing this measurement for several frequencies gives the user's hearing loss profile $h^*[j] j={0,1,\dots,N}$. 

In practice, this is measured for $N=7$ frequencies $(500Hz,1KHz,2KHz,3KHz,4KHz,6KHz,8KHz)$. This is too coarse a resolution to capture the exact hearing loss frequency response of a user as typically $N \ge 500$. 

So we cannot exploit these measurements ($Q_d$) alone to determine the personalizing filter $h^*$. However, these measurements do provide some information about the minimizer $h^*$. The audiogram test thus acts as the dimension query in \name.

\item[2.] \textbf{Query2 ($Q_f$): Satisfaction function sampling}: We can choose filters $h_j$ from the space of all filters $\mathcal{H} \subseteq \mathbf{R}^N$, apply it to any audio played to the user, and get their score $f(h_j)$. The user satisfaction function also has underlying sparsity because human hearing and hence audio perception is not the same across all frequencies \cite{pitch}. Under the perception function constraints, the audio personalization problem is well suited for sparsity-based Bayesian Optimization techniques like {\name}.
\end{itemize}

\ul{\textbf{Baselines}}:
\new

In Audio clinics, a user's audiogram is obtained through a coarse interpolation of these $Q_d$ pure tone measurements. We use this coarse audiogram as our \texttt{Baseline}

\ul{\textbf{Data Collection}}: 
\new

We recruit $3$ volunteers of $1$ male(s) and $2$ female(s) without hearing loss. To simulate hearing loss or distortion due to cheap speakers, We use two different corrupting filters $b_{1:2}$. 
\new

In the case of hearing loss, we use the publicly available hearing loss profiles in the NHANES \cite{nhanes} database as the corrupting filter $b1$. For random distortions (cheap speakers), we generate a random corrupting filter $b2$. 
\new

The audiogram and random distorting filter measurements are available at finite frequencies $(500Hz,1KHz,2KHz,3KHz,4KHz,6KHz,8KHz)$ and thus act as the $Q_d$. We have at most 7 $Q_d$ queries we can utilize. 
\new

A sample speech clip $a$ is filtered with the distorting filter $b1$ or $b2$ to obtain the \texttt{corrupted} clip This audio is what is heard by a person with hearing loss or as heard from a cheap speaker, $r = b_{1:2} * a$. 
\new

The goal of the audio personalization task is to apply different filter queries $Q_f$, $h_j$ to the clip $r$ to construct the user satisfaction function and optimize it to find $\hat{h}^*$, the personal filter. This filter when applied to the audio clip should make the resulting audio sound similar to the original uncorrupted speech clip (i.e.) $r * \hat{h}^* = \hat{a} \approx a$. 
\new

The personalization filter should counteract the distortion caused by $b1$ or $b2$. 